\documentclass[aps,showpacs,twocolumn]{revtex4}
\usepackage{amssymb,array,amsfonts,bm,dcolumn}
\usepackage[dvips]{graphicx}
\usepackage{amsmath}
\usepackage{color}
\usepackage{epsfig}

\begin{document}

\title{Quantum ratchet transport with minimal dispersion rate}

\author{Fei Zhan}
\author{S.~Denisov}
\author{A.~V. Ponomarev}
\author{P.~H\"anggi}
\affiliation{Institute of Physics, University of Augsburg,
Universit\"atsstr.~1, D-86135 Augsburg, Germany}

\begin{abstract}

We analyze the performance of quantum ratchets  by considering the dynamics of an initially localized wave packet loaded into a flashing periodic potential. The directed center-of-mass motion can be initiated  by the uniform modulation of the potential height, provided that the modulation protocol breaks  all relevant time- and spatial reflection symmetries. A poor performance of quantum ratchet transport is characterized by a slow net motion and a fast diffusive spreading of the wave packet, while the desirable optimal  performance is the contrary.  By invoking a quantum analog of the classical P\'eclet number, namely the quotient of the group velocity and the dispersion of the propagating wave packet, we calibrate the  transport properties of flashing quantum ratchets and discuss the mechanisms that yield low-dispersive directed transport.
\end{abstract}
\date{\today}

\maketitle
\section{INTRODUCTION}\label{sec:introduction}

Long-range, lasting and controllable quantum transport  in optical and magnetic potentials is a key to all-atom optical and atom-on-chip devices. The ratchet effect \cite{hanggi09,astumian, RGH,goychuk,epl,goychukLNP,ketz,quantum1,quantum2} presents a suitable toolbox for manipulations of ultracold matter, which  allows to set a quantum particle into directed motion without applying gradients or running-wave potentials \cite{running}. Periodic modulations of the confining potential is prerequisite for the effect to occur, and there are plenty of different setups and blueprints of ratchet machinery \cite{hanggi09,reimann}. An interesting class of ratchet systems are dissipation-free  Hamiltonian  ratchets, classical \cite{hanggi09,goychukLNP,flach}  and, even more intriguingly,  quantum ones \cite{hanggi09,ketz,goychukJPC,gong, leh, gong2007, pol2009, quantum1, quantum2,sergy07}.  For example, a bi-harmonic flashing potential with periodically and uniformly modulated potential height \cite{sergy07}, could be turned into a conveyer belt for a Bose-Einstein condensate (BEC) of rubidium atoms  \cite{salger}.

The  quantum ratchet effect can be utilized as a promising device capable to deliver  ultracold atoms to desirable locations. Yet the coherent  quantum transport in ac-driven periodic potentials  is severely hampered by  diffusion, additionally  enhanced by tunneling effects \cite{kolo94}. The issue of the transport efficiency \cite{eff1, eff2, eff3} becomes now of importance,  meaning that one should search for a set of optimal parameters in order to maximize the correspondingly chosen efficiency measure.  It is intuitive that the ratchet transport with large transport velocity and minimal dispersion rate would be preferable in the context of the delivery problem. This reasoning naturally leads to a quantum analog of the classical P\'{e}clet number concept \cite{pec}, which has already been used in the field of classical ratchets, both for overdamped \cite{lutz1, lutz2} and, as well, for  underdamped \cite{mach} models. However, the extension of this classical measure to the quantum limit is by no means straightforward. Different from the classical limit,  where the diffusive dispersion scales linearly, $\sigma^2(t) \sim t$,  the quantum diffusion is ballistic-like, $\sigma^2(t) \sim t^{2}$ \cite{kolo94}. This fact makes the direct implication of the P\'{e}clet number for the quantum ratchet problem not feasible. In this work we introduce a quantum analog of the P\'{e}clet number in order to specify the quality of ratchet transport. Using this concept we then aim in exploring the impact of  diffusion and tunneling on the  nonequilibrium quantum transport, focusing on  initial conditions in the form of a localized Gaussian-like wave packet.  Our objective is to identify  optimal regimes, when the directed transport is minimally swamped by the dispersive spreading. By use of the Floquet formalism \cite{grif}, which provides the complete information on the state of the system at any instant of time, we demonstrate how the analysis of the  Floquet spectrum of the driven system  allows one to judge  the quality of a flashing ratchet as a suitable delivery vehicle for ultracold atoms.

\begin{center}
\begin{table*}[t]
\begin{tabular}{|c|c|c|}
\hline
 &  &    \\
symmetry            &  transformation                   &   validity condition         \\
 &  &    \\
\hline
     &  &    \\
     &  &  $U(x)=U(-x)$  \\
$\mathcal{\hat{S}}_1$  &   $(x,p,t,\kappa)\rightarrow (-x,-p,t,-\kappa)$                          &     requires    \\
               &  & $s = 0$ or $\theta_p = \pm m\pi$, $m=0,1,...$ \\
 &  &    \\
\hline
 &  &    \\
     &    &     $E(t)=E(-t) $        \\
$\mathcal{\hat{S}}_2$   &           $(x,p,t,\kappa)\rightarrow (x,-p,-t,-\kappa) $                      &  requires       \\
                 &    & $E_2 = 0$ or $\theta = \pm m\pi$, $m=0,1,...$ \\
 &  &    \\
\hline
 &  &    \\
 &  &  $U(x+L/2)=-U(-x+L/2)$ and $E(t+T/2)=-E(t)$   \\
$\mathcal{\hat{S}}_3$   &  $(x,p,t,\kappa)\rightarrow (-x,-p,t+T/2,-\kappa)$                            &  requires \\
                 & &$s = 0$ (or $\theta_p = \pm m\pi/2$, $m=0,1,...$) and $E_2 = 0$ \\
 &  &    \\
\hline
 &  &    \\
     &    &  $E(t+t_s)=-E(-t+t_s)$ and $U(x)=-U(x+L/2)$           \\
$\mathcal{\hat{S}}_4$  &            $(x,p,t,\kappa)\rightarrow (x+L/2,-p,-t+2t_s,-\kappa)$                        &  where $t_s = 0$ or $T/2$, requires         \\
                 &    & $s =0$ and $E_2=0$ (or $\theta = \pm m\pi/2$, $m=0,1,...$)  \\
 &  &    \\
 \hline
\end{tabular}
\vspace{0.5cm} \caption{Symmetry transformations which reverse the current in the system (\ref{ham}-\ref{e}).} \label{tabl1}
\end{table*}
\end{center}

\section{Hamiltonian Quantum Ratchet Setup}\label{sec:model}
We consider a quantum particle loaded into a periodic potential, which is periodically modulated in time \cite{sergy07}. The dynamics of the system is governed by a Schr\"odinger equation,
\begin{equation}
i\hbar\frac{\partial}{\partial t}|\psi(t)\rangle=H(t)|\psi(t)\rangle,\label{schr}
\end{equation}
where the Hamiltonian $H(t)$ reads
\begin{equation}
 H(x,p,t)=\frac{p^2}{2m}+U(x)E(t).\label{ham}
\end{equation}
The flashing potential is formed by a bi-harmonic lattice of spatial period $L$, $U(x)=U(x+L)$,
\begin{equation}
 U(x)=K U_0 [\cos(kx)+s\cos(2kx+\theta_p)],~k=\frac{2\pi}{L};\label{u}
\end{equation}
whose amplitude is modulated by a bi-harmonic driving function,
\begin{align}
 E(t)\equiv E(t-t_0) =~~~~~~~~~~~~~~~~\notag\\ E_1\cos[\omega(t-t_0)]+E_2\cos[2\omega(t-t_0)+\theta], ~~~~\label{e}
\end{align}
of temporal period $T = 2\pi/\omega$. Here the time $t_0 \in [0, T]$ indicates the switch-on time of the driving force $E(t)$. We use  $L/2\pi$, $(m/k_L^2U_0)^{1/2}$, and $U_0$ as the units of distance, time and energy, correspondingly \cite{diss}. The system given by Eqs.~(\ref{ham}-\ref{e}) could describe the dynamics of a diluted cloud of ultracold atoms placed into optical potential formed by counter-propagating laser beams with periodically modulated intensities \cite{salger}.

The Hamiltonian \eqref{ham} is periodic both in time and space. The solution of the eigenvalue problem for the corresponding Floquet operator $U(t,t_0)$,  $|\psi(t+t_0)\rangle=U(t,t_0)|\psi(t_0)\rangle$,  provides the set  of eigenfunctions, $\{|\psi_{\alpha}(t)\rangle\}$.  The eigenfunctions satisfy the Floquet theorem, $|\psi_{\alpha,\kappa}(t)\rangle=e^{-i\varepsilon_{\alpha}[\kappa]t/\hbar}|\phi_{\alpha,\kappa}(t)\rangle$, $|\phi_{\alpha,\kappa}(t)\rangle=|\phi_{\alpha,\kappa}(t+T)\rangle$, and the Bloch theorem \cite{sergy07}. The Hilbert space of the system is sliced into invariant subspaces, each one of which is spanned by the states bearing the same quasimomentum value $\kappa \in [-\pi/L, \pi/L]$. The quasienergy values are confined to the interval $[-\hbar\pi/T,\hbar\pi/T]$, and quasienergies  of a given index $\alpha$ form a band across the $\kappa$-space, $\epsilon_{\alpha}[\kappa]$. For a given $\kappa$ and  theme $t$ the set of eigenstates $|\psi_{\alpha,\kappa}(t)\rangle$ forms a complete orthonormal basis for the corresponding subspace of the total system's Hilbert space. Any initial state, $|\psi(x,t=t_0)\rangle$, can be expanded in the Floquet basis, $|\psi(t=t_0)\rangle = \int_{-\pi/L}^{\pi/L} d\kappa \sum_{\alpha} C_{\alpha,\kappa}(t_0) |\phi_{\alpha, \kappa}  (t_0)\rangle$, and the time evolution of the state is  governed by \cite{ketz,sergy07}:
\begin{equation}\label{floquet}
|\psi(t)\rangle=\int_{-\pi/L}^{\pi/L}d\kappa\sum_{\alpha}C_{\alpha,\kappa}(t_0)
e^{-i\varepsilon_{\alpha,\kappa}t/\hbar}|\psi_{\alpha,\kappa}(t)\rangle.
\end{equation}

The asymptotic current generated by  the system, $J(t_0) = \lim_{t\rightarrow\infty} \frac{1}{t - t_0}\int_{t_0}^{t} \langle \psi(t')|\widehat{p}|\psi(t')\rangle dt'$, takes on a simple form in the Floquet basis \cite{sergy07}:
\begin{equation}\label{velocity}
J(t_0) = \int_{-1/2}^{1/2}d\kappa\sum_{\alpha}C_{\alpha,\kappa}(t_0)
\upsilon_{\alpha}[\kappa],
\end{equation}
where $\upsilon_{\alpha}[\kappa]$ is the the average velocity of the corresponding Floquet state, $\upsilon_{\alpha}[\kappa] = \langle\langle\phi_{\alpha \kappa}(t)|\widehat{p}|\phi_{\alpha \kappa}(t)\rangle\rangle_{T}$, and $\langle...\rangle_{T}$ denotes the averaging over one period of the driving. Following the Hellmann-Feynman theorem, the average velocity of an eigenstate is equal to the first derivative of the corresponding quasienergy band with respect to $\kappa$, $\upsilon_{\alpha}[\kappa] = \frac{1}{\hbar}d\epsilon_{\alpha} [\kappa]/d\kappa$ \cite{ketz,grif}.
The Hilbert space of the system can be imagined as a set of ``conveyer belt" -- eigenstates, each one of which is moving with its own velocity, and the overall ratchet current depends on how  the initial wave packet $|\psi(t=t_0)\rangle$ was distributed among the various  belts \cite{sergy07,salger}.

\section{SYMMETRIES IN QUASIMOMENTUM SPACE}\label{sec:sym}
The transport properties of Floquet states are governed by the space-time (a)symmetries of corresponding Hamiltonian \cite{sergy07}.  There are four types of relevant  symmetry transformations for the Hamiltonians of the type \eqref{ham}, see Tab.~1. Two of them, $\mathcal{\hat{S}}_1$ and $\mathcal{\hat{S}}_3$, demand the space inversion, and the remaining  two, i.e., $\mathcal{\hat{S}}_2$ and $\mathcal{\hat{S}}_4$, involve the explicit time reversal. Each transformation changes the sign of the quasimomentum, and maps every Floquet band onto itself, a negative branch onto a positive one, and vice versa, $\hat{S}_{i}:\varepsilon_{\alpha}[\kappa]=\varepsilon_{\alpha}[-\kappa]$. The Floquet states with $\kappa \neq 0$ may possess nonzero velocities, $\upsilon_{\alpha}[\kappa]$, even in the presence of a symmetry; but in this case we have  $\upsilon_{\alpha}[-\kappa] = - \upsilon_{\alpha}[\kappa]$.  Note that only the eigenstates from the center of the Brillouin zone, $\upsilon_{\alpha}[0]$, are mapped onto itself under the transformations from Tab.~1. The symmetry analysis of quantum ratchets performed in previous works was restricted  on  transport properties involving the zero-quaismomentum states only \cite{sergy07}. The main result of the analysis was that whenever one of the symmetries listed in Tab.~1  holds, all the Floquet states with $\kappa=0$ are nontransporting, i.e. $\upsilon_{\alpha}[0]=0$. Independent of a particular shape of the initial cloud and the choice of starting time $t_0$, the asymptotic current would be absent --  for any initial state drawn from the subspace $\kappa = 0$.

Although such the initial state can serve as a good approximation of a diluted BEC cloud  smeared over sufficiently many wells of the periodic potentials \cite{cloud},  it cannot mimic the experimentally relevant situation with a partially delocalized wave packet, $\psi(x,t_0)$. Such a more realistic wave packet occupies a certain  region in the $\kappa$-space, and therefore, Floquet states of different quasimomenta $\kappa$ and band indices $\alpha$, will contribute to the wave-packet dynamics according to the corresponding   weights, $C_{\alpha, \kappa}(t_0)$, see Eq. \eqref{floquet}.

We could make the reasonable assumption  that at  time $t_0$ we start  out with the initial cloud in form of a  symmetric wave-packet, $\psi(-\kappa,t_0) = \psi(\kappa,t_0)$. As it is the case with the  dynamics restricted to the subspace of $\kappa = 0$, the evolution of a wave packet depends on the initial starting time $t_0$, Fig. 1(a). In contrast to its classical counterpart, however, where the chaotic dynamics erases the memory about the  initial conditions of the system preparation exponentially fast  \cite{sergy07, salger}, a fully {\it coherent} quantum ratchet maintains this information, which is encoded in the coefficients $C(t_0)$ forever. All the symmetries from Table 1, except the first one, $\hat{S_1}$, involve time transformations  -- time inversion in $\hat{S}_2$, time shift in $\hat{S}_3$, and a combined transformation in $\hat{S}_4$ -- so that the contributions from the eigenstates with opposite quasimomenta, might be different for a given, fixed $t_0$, $C_{\alpha,-\kappa}(t_0) \neq C_{\alpha,\kappa}(t_0)$, even when one (or all) of the symmetries $\hat{S}_2, \hat{S}_3, \hat{S}_4$ are obeyed. The asymptotic current $J(t_0)$, see Eq. \eqref{velocity},  may acquire then a nonzero value. Only the presence of the symmetry  $\hat{S}_1$  guarantees the  absence of the quantum ratchet  current for any symmetric chosen initial wave packet and for any choice of the starting time $t_0$.

The averaging over $t_0$ provides the way for the realization of the symmetries $\hat{S}_2, \hat{S}_3$ and $\hat{S}_4$, so that the equalities for the \textit{averaged} contributions, $\langle C_{\alpha, -\kappa}(t_0) \rangle_{t_0} = \langle C_{\alpha, \kappa}(t_0) \rangle_{t_0}$, now hold whenever at least one of these symmetries is present. The contributions with opposite velocities cancel each other out, and the $t_0$-averaged current
\begin{equation}
\bar{J} = \langle J(t_0)\rangle_{t_0}
\end{equation}
 then equals  zero, see Fig.1(b).
%
%

\begin{figure}[t]
\includegraphics[width=0.49\textwidth]{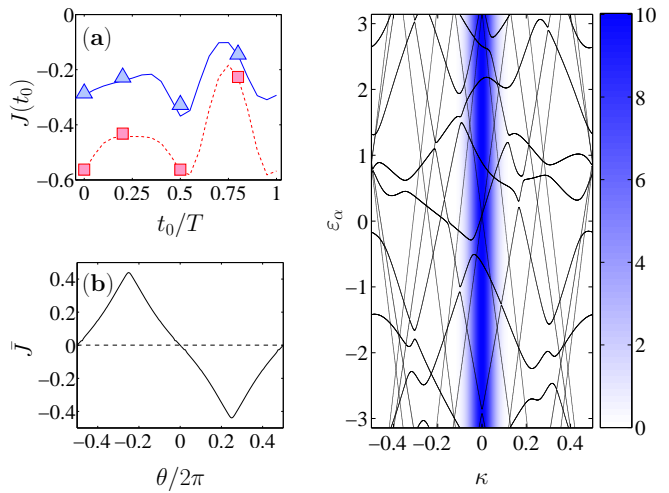}
\caption{(Color online) \label{fig:curtheta} Left panel: (a) Quantum ratchet current as the function of the starting time $t_0$. The  lines correspond to the asymptotic current  calculated by using Eq.~\eqref{velocity} for $\theta=\pi/2$ and $\theta_p=2.2$ (solid blue); $\theta_p=-\pi/2$ (dashed pink).  Triangles (squares) correspond to the current calculated by the direct propagation of an initial wave packet up to the time $100T$. (b) The averaged current $\bar{J} = \langle J(t_0)\rangle_{t_0}$ at the spatial phase difference $\theta_p=\pi/2$ versus the phase difference $\theta$ between the two harmonic driving components.  Right panel: The  quasienergy spectrum  as a function of the quasimomentum. The color plot shows  the $\kappa$-representation of the   initial wave packet, Eq.~\eqref{initial}, with $\sigma_0=4\pi$. The other parameters are  $\omega=1$, $\hbar=1$, $E_1=0.9$, $E_2=0.45$, $K=0.9$, $s=0.7$, $\theta=\pi/2$, $\theta_p=-\pi/2$. The current is in units of the recoil momentum.}
\end{figure}

\section{Quality measure of quantum ratchet transport}\label{sec:efficiency}

In order to explore the transport performance of Hamiltonian quantum ratchets we use here two independent  approaches. Namely, we employ the direct integration of the Schr\"odinger equation \eqref{schr} on the interval $[-100 \times L,100 \times L]$ by using a grid with $10^{4}$ points,  and, independently,  use the Floquet approach based on the plane-wave expansion \cite{sergy07}. As our initial wave function we used a Gaussian wave-packet, reading,
\begin{equation}
 \psi(x,t_0=0)=(2\pi\sigma_0^2)^{-1/4}\exp\left(-\frac{x^2}{4\sigma_0^2}\right),\label{initial}
\end{equation}
with the variance $\sigma_0=4\pi$, Fig.~\ref{fig:curtheta}, right panel.

The results of both numerical schemes exhibit very good agreement, see Fig.~\ref{fig:curtheta}(a). The average current $\bar{J}(\theta)$ matches the prediction of the symmetry analysis, revealing the property $\bar{J}(-\theta)=-\bar{J}(\theta)=\bar{J}(\theta + \pi)$ \cite{sergy07},
see in Fig.~\ref{fig:curtheta}(b).

A desirable performance of a quantum ratchet transport is  when the spreading of the wave packet constitutes a slow process when compared to the  overall directed motion of the packet center of mass. This limit would correspond to a highly coherent transport, when the ballistic motion of the atomic cloud is minimally hampered by  diffusion. The natural measure of the quality of the ratchet transport would  then be the quotient of the directed current and a proper diffusive characteristics. The  diffusion is naturally quantified by using the dispersion of the wave packet, $\sigma^2(t)=\langle x(t)^{2} \rangle - \langle x(t) \rangle^{2}$. In the case of normal diffusion, dispersion scales like $\sigma^2(t)\propto t$, and in the classical limit, it perfectly describes the kinetics of over- \cite{lutz1,lutz2,eff3} and under-damped \cite{mach} ratchets. Yet the situation with normal diffusion is rarely the case even in the classical, dissipation-free limit, where  ac-driven Hamiltonian systems demonstrate typically superdiffusive kinetics, $\sigma^2(t) \propto t^{\gamma}$, with a parameter-sensitive exponent $1 < \gamma < 2$  \cite{anomal1, anomal2}. In the coherent quantum limit the diffusion is uniformly ballistic in the asymptotic limit $t \geq t_{\text{bal}}$,  $\sigma^2(t) \propto t^{2}$, with a parameter-sensitive transient time scale $t_{\text{bal}}$ \cite{kolo94}, see the inset in Fig.~2(a).
Therefore, the P\'{e}clet number, $Pe = J L/D$ \cite{pec}, which previously was successfully applied for the calibration of classical ratchets,  presents now an inappropriate measure in the quantum context. We next introduce here its quantum analog,  named  \textit{quality of  ratchet transport},
\begin{equation}
Q = Q(t_0) =  J^{2}(t_0)/D_{\text{eff}}(t_0)\;,\label{peclet}
\end{equation}
where $D_{\text{eff}} = \lim_{t\rightarrow \infty}\sigma^2(t)/t^{2}$. This so introduced ``quality'' quantifier,  as it is the case  for the classical predecessor, $Pe$,   is dimensionless.

\begin{figure}[t]
{\center\includegraphics[width=8cm]{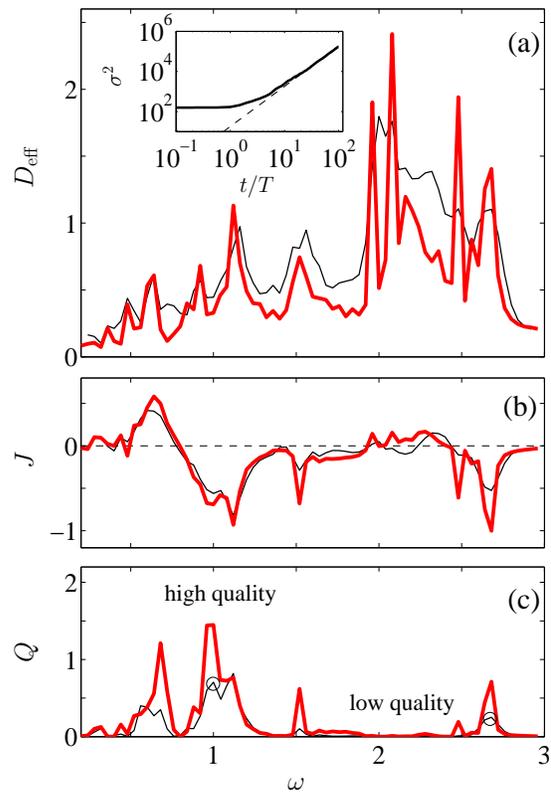}}
\caption{\label{fig:cdp}(Color online) (a) Dependence of the effective diffusion coefficient $D_{\rm eff} = D_{\rm eff}(t_0 = 0)$ on the driving frequency $\omega$ for the initial wave packets, Eq. ~\ref{initial}, of the dispersion  $\sigma_0=4\pi$ (thin black line) and $\sigma_0=40\pi$ (thick red line). The inset depicts the time evolution of the  dispersion $\sigma^2(t)$ of the wave packet with $\sigma_0=4\pi$  for  $\omega=1$. The dashed line corresponds to the asymptotic limit $\sigma^2(t) \propto t^2$. (b) The quantum ratchet current $J(t_0 = 0)$ and (c) the  quality of the coherently directed quantum transport $Q(t_0 = 0)$ as  functions of the driving frequency $\omega$. The two circled values at $\omega = 1$ and at $\omega = 2.68$ are discussed further as  functions of quasienergy-quasimomentum characteristics in Figure 3. The diffusion coefficient and the current are in units of the square of the recoil momentum and the recoil momentum, correspondingly.}
\end{figure}

Among the several parameters of the system dynamics it is  the driving frequency $\omega$ which presents a most suitable control parameter. This driving frequency $\omega$ allows one to tune the ratchet into resonant regimes  \cite{sergy07}. The frequency of the driving can conveniently be varied in experiments \cite{salger}. The dependence of the effective diffusion coefficient $D_{\text{eff}}$ on the driving frequency, see Fig.~2(a),  does not keep abreast with the corresponding current dependence, see Fig.~2(b) (thin line), therefore producing regions of different transport quality. Note that although the asymptotic absolute current values $|\bar J|$ are identical at the driving values $\omega=1$ and $\omega=2.68$, the quality of the transport in the first point is almost three times better than in the second.

In order to gain insight into mechanisms of highly-coherent quantum transport, we resort to the analysis of the system Floquet spectrum. First we sorted out all the Floquet bands according to their kinetic energy, and then ordered them into an ascending order. This is a reasonable procedure, since the states with kinetic energies that are much larger than the characteristic energy $K U_0$, would not contribute tangibly to the initially prepared  wave packet. Finally we left with a small number of relevant Floquet bands of low energy, see Figs.~3(a,b).

The relative slopes of the energy bands determine not only the velocity of Floquet eigenstates, but also the diffusion properties of the system. By using the results of Ref.~\cite{ketz}, we can rewrite the expression for the effective diffusion coefficient in the following form (the detailed derivation  is given in the Appendix),
\begin{align}
D_{\text{eff}}(t_0)=\frac{1}{2}\iint_{-1/2}^{1/2} d\kappa_1d\kappa_2&\sum_{\alpha\beta}|C_{\alpha,\kappa_1}(t_0)|^2|C_{\beta,\kappa_2}(t_0)|^2\notag\\
&\times\left(\frac{d\varepsilon_{\alpha}[\kappa_1]}{d\kappa_1}-\frac{d\varepsilon_{\beta}[\kappa_2]}{d\kappa_2}\right)^2.~~~~~\label{diff}
\end{align}

\begin{figure}[t]
{\center\includegraphics[width=8cm]{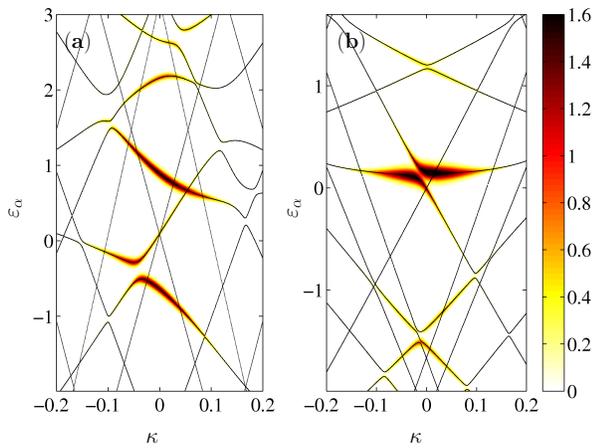}}
\caption{\label{fig:fig2}(Color online) The quasienergy bands $\varepsilon_\alpha$ as a function of quasimomentum $\kappa$. (a) The case of  high-quality quantum ratchet transport realized at the circled value $\omega=1$ in Fig.~2(b). (b) The case of a low-quality quantum ratchet transport as realized at the circled value $\omega=2.68$ on Fig.~2(b). The depicted ``widths'' relate the color coding [from dark (maximal) to bright (minimal)] of a band encode to the corresponding weight, $|C_{\alpha,\kappa}(t_0 = 0)|^{2}$. The weights were obtained by projecting the initial Gaussian wave packet \eqref{initial} onto the corresponding Floquet state $\phi_{\alpha,\kappa}(t_0 = 0)$.}
\end{figure}

With the Eqs.~\eqref{velocity} and \eqref{diff} in hands, we can discuss  two extreme limits.
Consider first the ultimate case with only two bands, $i$ and $j$, that are populated initially. If the bands are parallel at the vicinity of $\kappa = 0$, both have non-zero slopes and share the initial cloud equally, $|C_{i}(t_0)|^{2} = |C_{j}(t_0)|^{2}=1/2$; then the corresponding effective diffusion coefficient would be equal to zero. Accordingly, the quality $Q$ would be infinite in this case. This regime is approached at the frequency $\omega=1$ (point of a high-quality transport in Fig.~2(c)), see Fig.~3(a).

The situation at the frequency $\omega=2.68$ is contrary to the discussed one, see Fig.~3(b). There we also find  only two mostly populated bands which exhibit an avoided crossing (AC). The bands have different slopes, and according to the last factor in Eq.~\eqref{diff}, this implies  strong diffusion. Both bands are almost horizontal within the AC region, so that they yield almost no current. Most of the contribution to the overall current stems from the  parts of the bands, located away from the AC region. Although  the bands have a large slope, thus producing a directed  current which matches  that of the regime at the point $\omega=1$, the presence of strong diffusion decreases the transport quality by factor of three (point of a low-quality transport in Fig.~2(c)).

\section{CONCLUSION}

In summary, we have investigated the coherent quantum transport of a quantum wave packet in a flashing periodic potential. The regime of high-quality ratchet transport  corresponds to  the ballistic center-of-mass motion  minimally obstructed by quantum diffusion. The quality is a tunable characteristics, and it is indeed possible to optimize the quantum ratchet performance by tailoring the parameters of the driving flashing potential. The transport quality $Q$ is also an experimentally relevant characteristics: the ratchet current can be measured by using the time-of-flight technique \cite{salger}, while the dispersion of BEC cloud can be estimated directly from absorption images \cite{absorb}.

The quality of the transport depends on the initial time $t_0$, since the band  populations, $|C_{\alpha,\kappa}(t_0)|^2$, are functions of starting time $t_0$. Therefore,  different choices of the starting time $t_0$ may lead to regimes of different transport quality even when all the driving parameters of the system Hamiltonian are fixed. The dependence on the dispersion of the initial wave packet, $\sigma_0$ \eqref{initial},  is even more stringent. Namely, the localization in the $x$-space is inversely proportional
to the localization in $\kappa$-space. Therefore, a very {\it delocalized} initial wave packet, smeared over many periods of the optical potential, would be strongly {\it localized} at the point $\kappa=0$ in the quasimomentum space.
The asymptotic transport dynamics of the wave packet can  be evaluated  from the behavior of the quasienergy bands at the  center of the Brillouin zone, by using  Eqs.~\eqref{velocity} and \eqref{diff}. The localization in $\kappa$-space, thereby, makes the effect sharper and typically leads to the enhancement of the quality of ratchet transport, see Fig.~2(c).
On the contrary, the ratchet effect would be faint for an initial wave packet in the form of a very narrow peak in $x$-space. Strong localization of the cloud would correspond to an almost uniform distribution over the  Brillouin zone, and since the average band velocities are all equal to zero,  $\int_{-\pi/L}^{\pi/L} \upsilon_{\alpha}[\kappa] dk = \epsilon_{\alpha} [-\pi/L] - \epsilon_{\alpha} [\pi/L] = 0$, such the initial distribution  will yield almost imperceptible current. At the same time the spreading rate would be huge. Paradoxically enough, a very narrow wave packet will lead to quantum ratchet transport of a very poor quality \cite{motor}.

Our results can be considered as a first step towards an optimization scheme of cold atom engines. The ultimate aim of the research in this direction  is an operational recipe, a protocol,  which would allow one to maximize the quality of a quantum ratchet transport  without going deeply into the analysis of the particular system of interest. One of the intriguing subdirections is the role of disorder in the functioning  of quantum ratchets. Namely, a following question is of special interest: Can a pinch of disorder in the underlying optical potential improve the performance of quantum ratchets? In other words, can the disorder-induced localization \cite{diss}  slow down the diffusive spreading without affecting the center-of-mass velocity?
\\
\appendix
\section{EFFECTIVE DIFFUSION COEFFICIENT}
We start out from the expression for the mean square dispersion given in the Appendix~B of Ref.~\cite{ketz},
\begin{align}
\sigma^2 = t^2\int_{-1/2}^{1/2}d\kappa\sum_{\alpha}|C_{\alpha\kappa}(t_0)|^2\left(\frac{d\varepsilon_{\alpha}[\kappa]}{d\kappa}\right)^2\notag\\
-\left(t\int_{-1/2}^{1/2}d\kappa\sum_{\alpha}|C_{\alpha\kappa}(t_0)|^2\frac{d\varepsilon_{\alpha}[\kappa]}{d\kappa}\right)^2+O(t).
\end{align}
Here and in the following we drop the dependence on the initial starting time $t_0$ for the results on l.h.s. of equations.
The corresponding effective diffusion coefficient is given by,
\begin{align}
D_{\text{eff}}=&\int_{-1/2}^{1/2}d\kappa\sum_{\alpha}|C_{\alpha\kappa}(t_0)|^2\left(\frac{d\varepsilon_{\alpha}[\kappa]}{d\kappa}\right)^2\notag\\
&-\left(\int_{-1/2}^{1/2}d\kappa\sum_{\alpha}|C_{\alpha\kappa}(t_0)|^2\frac{d\varepsilon_{\alpha}[\kappa]}{d\kappa}\right)^2.\label{deffapp}
\end{align}
We consider the integral with respect to $\kappa$ as the limit of the summation,
\begin{equation}
\int_{-1/2}^{1/2}d\kappa\rightarrow\lim_{\Delta\kappa\rightarrow0}\sum_{\kappa=0}^{n}\Delta\kappa,
\end{equation}
with the condition $n\Delta\kappa=1$. Next, the expression \eqref{deffapp} is transformed into a summation with respect to  indices $\kappa$ and $\alpha$. We reorder all the terms from $1$ to $N$, and recast Eq.~\eqref{deffapp} in the following form,
\begin{equation}
D_{\text{eff}}=\sum_{i=1}^{N}a_ix_i^2-\left(\sum_{i=1}^{N}a_ix_i\right)^2,\label{deffax}
\end{equation}
where $a_i=\Delta \kappa|C_{\alpha\kappa}(t_0)|^2$ and $x_i=d\varepsilon_{\alpha}[\kappa]/d{\kappa}$.

Due to the norm preservation, we obtain
$\sum_{i=1}^{N}a_i=1$.
We further simplify Eq.~\eqref{deffax}, and by using a simple algebra, arrive at
\begin{align}
D_{\text{eff}}=&\sum_{i=1}^{N}a_ix_i^2-\sum_{i,j=1}^{N}a_ia_jx_ix_j\notag\\
=&\sum_{i=1}^{N}a_i(1-a_i)x_i^2-2\sum_{\substack{i=1\\j>i}}^{N}a_ia_jx_ix_j\label{ax1}\\
=&\frac{1}{2}\sum_{i,j=1}^{N}a_ia_j(x_i-x_j)^2. \label{ax5}
\end{align}
We next return  to the summation with respect to the indices $\kappa$ and $\alpha$ and take the limit  $\Delta\kappa \rightarrow 0$. The integral form is recovered and therefore we end up with Eq.~\eqref{diff}.


\begin{thebibliography}{99}
\bibitem{cloud} The narrow `slice' in the quasimomentum space can be prepeared by using the Bragg selection techniques. See in Ref. [17] for more detail.
\end{thebibliography}
\end{document}